%% file: main.tex
\documentclass{vldb}
\usepackage[utf8]{inputenc}
\usepackage{verbatim}
\usepackage{url}
\usepackage{graphicx}
\usepackage{caption}
\usepackage[tight,TABTOPCAP]{subfigure}
\usepackage{wrapfig}
\usepackage{color}
\usepackage{xspace}
\usepackage{charter}
\usepackage{bbm}
\usepackage{amsmath}
\DeclareMathOperator*{\argmin}{argmin}
\usepackage{balance}
\usepackage{soul}

\newcommand{\nom}[1]{\texttt{nominal}\xspace}
\newcommand{\ch}[1]{\texttt{Chao92}\texttt{nominal}\xspace}

\newtheorem{problem}{Problem}

\newtheorem{example}{Example}

\newcommand{\naive}{na\"{\i}ve\xspace}

\title{A Data Quality Metric (DQM)}
\subtitle{How to Estimate the Number of Undetected Errors in Data Sets}

\author{
\begin{tabular}{ccc}\\
Yeounoh Chung\textsuperscript{1} & Sanjay Krishnan\textsuperscript{2} & Tim Kraska\textsuperscript{1}  \\
\end{tabular}\\
\begin{tabular}{c}\\
\textsuperscript{1}\eaddfnt{Brown University,} 
\eaddfnt{firstname\_lastname@brown.edu}\\
\textsuperscript{2}\eaddfnt{UC Berkeley,}
\eaddfnt{firstname@eecs.berkeley.edu}
\end{tabular}
}

\date{February 2017}

\begin{document}

\maketitle

\begin{abstract}
Data cleaning, whether manual or algorithmic, is rarely perfect leaving a dataset with an unknown number of false positives and false negatives after cleaning. In many scenarios, quantifying the number of remaining errors is challenging because our data integrity rules themselves may be incomplete, or the available gold-standard datasets may be too small to extrapolate. As  the  use  of  inherently fallible crowds becomes more prevalent in data cleaning problems, it is important to have estimators to quantify the extent of such errors. We propose novel species estimators to estimate the number of distinct remaining errors in a dataset after it has been cleaned by a set of crowd workers -- essentially, quantifying the utility of hiring additional workers to clean the dataset. This problem requires new estimators that are robust to false positives and false negatives, and we empirically show on three real-world datasets  that existing species estimators are unstable for this problem, while our proposed techniques quickly converge.
\end{abstract}

\input{introduction}

\input{background}

\input{species-estimate}

\input{estimators-fp}

\input{priotization}
\input{experiment}
\input{related}

\input{conclusion}

\bibliographystyle{abbrv}
\fontsize{8.0pt}{9.0pt} \selectfont
\bibliography{main}

\balance

\end{document}

%% file: introduction.tex
\section{Introduction}
\label{sec:intro}

 Data scientists report that data cleaning, including resolving duplicates and fixing inconsistencies, is one of the most time-consuming steps when starting a new project~\cite{nytimes, DBLP:conf/sigmod/ChuIKW16}. A persistent challenge in data cleaning is coping with the ``long tail'' of errors, which can affect algorithmic, manual, and crowdsourced cleaning techniques. For example, data integrity rules can be incomplete and miss rare problems, and likewise, crowd workers that are not familiar with a particular domain might miss subtle issues. Consequently, even after spending significant time and/or money to clean and prepare a dataset, there may still be a large number of undetected errors.

This paper explores whether it is possible to estimate the number of errors that remain in a dataset purely from observing the results of previous cleaning operations (i.e., without rules or gold standard data). By quantifying errors, we mean that if the observed data cleaning method was given infinite resources (time, money, or workers), how many more erroneous records can one expect to find that are currently not marked. Being able to answer such a question based on the results of previous data cleaning operations is valuable because, in practice, multiple fallible data cleaning techniques or crowd-workers are employed in the hope of increasing error detection coverage without quantitative guidance.

\subsection{Quantifying the Remaining Errors}
\begin{figure}[t!]
 \centering
 \includegraphics[width=\columnwidth]{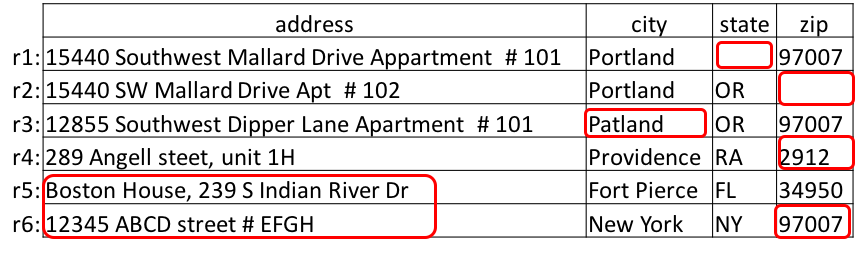}
  \caption{Erroneous US home addresses: $r1$ and $r2$ contain missing values; $r3$ and $r4$ contain invalid city names and zip codes; $r1$, $r3$, and $r6$ violate a functional dependency ($zip~\rightarrow~city,~state$);
  $r5$ is not a home address, and $r6$ is a fake home address in a valid format.} 
  \label{fig:mal_addr}\vspace{-3mm}
\end{figure}

While this is a seemingly simple question, it is actually extremely challenging to define data quality without knowing the ground truth  \cite{DBLP:journals/cacm/PipinoLW02, DBLP:journals/jdiq/CheahP15, DBLP:journals/jdiq/EvenS09,DBLP:journals/jdiq/SessionsV09, DBLP:journals/tkde/FanGLX11,DBLP:journals/sigmetrics/KeetonMW09}; previous works define data quality through counting the losses to gold standard data or violations of the constraint rules set forth by domain-specific heuristics and experts \cite{yakout2011guided,bohannon2005cost,chomicki2005minimal,cong2007improving,lopatenko2007efficient}. In practice, however, such ground truth data or rules are not readily available and are incomplete (i.e., there exists a ``long tail'' of errors). 
For instance, take a simple data cleaning task where we want to identify (and manually fix) malformed US home addresses in the database, shown in Figure~\ref{fig:mal_addr}.
As in Guided Data Repair (GDR)~\cite{yakout2011guided}, we might have a set of rules to propose repairs for missing values ($r1$, $r2$) and functional dependency ($r1$, $r3$, $r6$), but not for US state/city name misspellings ($r3$, $r4$) or wrong home addresses ($r5$, $r6$), in a seemingly valid format that only the most observant crowd-workers might catch. Once errors are identified, a human can verify the proposed errors and automatic repairs. Similarly, as in CrowdER~\cite{wang2012crowder}, we can run inexpensive heuristics to identify errors and ask crowd-workers to confirm. In both of these cases, the fallibility of the system in the form of false negative (e.g., ``long tail'' or missed errors) and false positive (e.g., even humans can make mistakes) errors is a big concern.

In this work, we set out to design a statistical estimator to address both of the issues. That is, we need to estimate the number of remaining errors without knowing the ground truth in the presence of false negative and false positive errors. 
A simple approach is to extrapolate the number of errors from a small \emph{``perfectly clean''} sample (as in~\cite{wang1999sample}): (1) we take a small sample, (2) perfectly clean it manually or with the crowd, and (3) extrapolate our findings to the entire data set. 
For example, if we found 10 new errors in a sample of 1000 records out of 1M records, we would assume that the total data set contains 10000 additional errors. However, this \naive approach presents a {\em chicken-and-egg} paradox. 
If we clean a very small sample of data, it may not be representative and thus will give an inaccurate extrapolation or estimates based off it.
For larger samples, how can the analyst know that the sample itself is perfectly clean without a quality metric?

To this end, we design a statistical estimator based on the principle of diminishing returns, which basically states that every additional error is more difficult to detect.
For example, with experts or crowd sourcing, the first (crowd-)worker to pass over the data set finds more new errors, than every subsequent worker, and so on. 
The key insight is to estimate this diminishing return rate (i.e., fewer errors are found in each pass) and project this rate forward to estimate the number of errors if there were an infinite number of workers.

Note that, while we use crowd-sourcing as our motivation, the same observation also applies for purely algorithmic cleaning techniques \cite{getoor2012entity} (e.g., machine learned error classifiers). 
Each additional data cleaning algorithm applied to a data set will have a diminishing utility to the analyst. 
However, in this paper, we focus on crowd-sourced cleaning techniques because they empirically achieve high quality results and are widely used in the industry \cite{marcus2015crowdsourced}. 
Exploring pure algorithmic techniques, which make some problems easier but others harder as explained later, are beyond the scope of this paper. 
In this work, we present examples on entity resolution and malformed entities, although our statistical techniques can also be applied to other types of data errors as long as the errors are countable (i.e., independent workers provide the amount of dirty data or errors they could find in the dataset).

\subsection{Background: Crowd-Based Cleaning}

Most practical data cleaning tools rely on humans to verify the errors or repairs~\cite{abedjan2016detecting}, and  
Crowd-based cleaning techniques use humans to identify errors in conjunction with algorithmic pre-processing. 
To overcome human error, the prevailing wisdom is to hire a sufficient number of workers to guarantee that a selected set of records is reviewed by multiple workers.
This redundancy helps correct for false positives and false negatives, and there are many well-studied techniques to do so (e.g., Expectation Maximization and Voting~\cite{DBLP:conf/nips/ZhangCZJ14, DBLP:conf/aaaiss/LangeL12}).
The core idea of all these techniques is that with more and more workers, the majority of the workers will eventually agree on the ground truth.

Consider a two-stage entity resolution algorithm as proposed in CrowdER~\cite{wang2012crowder}. 
During the first stage, an algorithmic similarity-based technique determines pairs that are likely matches (potential errors), unlikely matches, and critical candidates for crowd verification. 
For example, likely matches could be the records pairs with a Jaccard distance of 1, unlikely matches as the ones with a similarity below 0.5, and  everything in between as uncertain (i.e., the candidate matches).  
During the second stage, a crowd-worker is assigned to one of the candidate matches and determines if the pair is a match or not. 
However, assigning a pair to a single worker is risky since he or she can make a mistake.
To improve the quality, we could assign a second and third worker to check the same items again.
Although the quality of the cleaned data set improves, the impact of every additional worker decreases.

In this work, we also assign multiple workers to each item to verify and fix the errors, but at random. This enables our technique to leverage the diminishing utility of workers for the estimation. In other words, instead of assigning a fixed number of workers (e.g., three to form a quorum) to all items, we distribute a small subset of the dataset to each worker uniformly at random.
Such redundancy in worker assignment gives rise to a reliable quality metric based on the number of remaining error estimations; however, it seems wasteful from the data cleaning perspective. In Section~\ref{sec:exp}, we empirically show that the added redundancy, due to the random assignment, is marginal compared to the fixed assignment (exactly three votes per item).

\subsection{Our Goal and Approach}

{\bf Our goal is to estimate the number of all (eventually) detectable errors}. 
We are not concerned with the errors that are not detectable even with infinite resources. 
In the case of our two-stage data cleaning with crowd-sourcing, infinite resources means an infinite number of crowd-workers, who would eventually reach the correct consensus decisions (i.e., dirty or clean) for all  candidate matches. 

Maybe surprisingly, this problem is related to estimating the completeness of query results using species estimation techniques as first proposed in \cite{DBLP:journals/cacm/TrushkowskyKS16}.
We can think of our data quality problem as estimating the size of the set containing all distinct errors that we would discover upon adding more workers.
Each worker marks a subset of records as dirty, with some overlaps with other workers.
The idea is to estimate the number of distinct records that will be marked erroneous if an infinite number of workers,  $K \rightarrow \infty$, are added.

Unfortunately, existing species estimation techniques  \cite{DBLP:journals/cacm/TrushkowskyKS16,unknownunknowns} to estimate the completeness of a set, do not consider that workers can make mistakes.
At the same time in any real data cleaning scenario, workers can make both false negative errors (a worker fails to identify a true error) and false positive errors (a worker misclassifies a clean item as dirty). 
It turns out that false positives have a profound impact on the estimation quality of how many errors the data set contains. 
This is because species estimators rely on the number of observed  ``rare'' items as a proxy for the number of remaining species, and this number can be highly sensitive to a small number of false positive errors.

\noindent\textbf{Contributions: }
To address this issue, we reformulate the estimation problem to estimate the number of distinct changes in the majority consensus. 
To the best of our knowledge, this is the first research done on the diminishing return effect, in the context of consensus, and to develop techniques to estimate the number of remaining errors in a data set, without knowing the true number of errors in the original dataset or a complete set of rules to define a perfect dataset.

In summary, we make the following contributions:
\begin{itemize}
    \item We formalize the link between data quality metrics and species estimators, which enables estimation of the number of remaining errors without knowing the ground truth in advance.
    \item Analytically and empirically, we show that traditional species estimators are very sensitive to errors from the crowd. 
    \item We propose a variant of the species estimation problem that estimates the number of distinct majority switching events, find that this new estimator is more accurate in the presence of noise (e.g., misclassified errors), and show how this estimate can be used to determine the quality of the data set.
    \item We evaluate our switch-based quality metric using real- and synthetic data sets and find that they provide much more accurate and reliable estimates with fallible crowds. 
\end{itemize}

The paper is organized as follows: Section \ref{sec:setup} formalizes the data quality metric problem and discusses how \naive baseline approaches fall short. Section \ref{sec:nofp} describes species estimation solutions to the data quality metric problem. Section \ref{sec:fp} presents a re-formulation of the original problem and proposes a solution that is robust to false positive errors. Section \ref{sec:priority} describes how to non-uniformly sample records. The remaining sections describe an evaluation on real and simulated data sets.

%% file: background.tex
\section{Problem Setup And Baselines}\label{sec:setup}
In this section, we formalize our assumptions about the cleaning process and the problem setup. 

\subsection{Basic Setup}
Every data set $R$ consists of a set of records $\{r_1,...,r_N\}$.
Some subset of $R$ is erroneous $R_{dirty} \subseteq R$. 
Let $W=\{w_1, ..., \\w_K\}$ be the set of \emph{``fallible''} workers, and each worker $w_i \in W$ observes a subset of records $S^{(i)} \subseteq R$ and identifies those records that are dirty $S^{(i)}_{dirty}$ and clean $S^{(i)}_{clean}$, with some unknown error rates. Workers might have different error rates as well as a different set of internal rules to identify errors, but a sufficient number of workers would correctly identify errors in  consensus~\cite{DBLP:conf/nips/ZhangCZJ14,DBLP:conf/aaaiss/LangeL12}.  

Therefore, we have a collection of sets $\mathcal{S}_{dirty} = \{S^{(1)}_{dirty},..., \linebreak S^{(K)}_{dirty}\}$ of all the records that the crowd identified as dirty and a collection of sets of all the records that the crowd identified as clean $\mathcal{S}_{clean} = \{S^{(1)}_{clean},...,S^{(K)}_{clean}\}$.
The worker responses can be concisely represented in a $N \times K$ matrix $\mathcal{I}$, where every column represents the answers from a worker $k$, and which entries are $\{1,0,\emptyset\}$ denoting dirty, clean, unseen respectively. 

\begin{problem}[Data Quality Estimation]
Let $\mathcal{I}$ be a $N \times K$ worker-response matrix with entries $\{1,0,\emptyset\}$ denoting dirty clean, unseen respectively.
The data quality estimation problem is to estimate $|R_{dirty}|$ given $\mathcal{I}$.
\label{problem:random}
\end{problem}

This model is general enough to handle a variety of crowd-sourced cleaning operations such as missing value filling, fixing common inconsistencies, and removing irrelevant or corrupted records; the errors only need to be identifiable and countable by some of the fallible workers in $W$. Notice that the model does not require the ground truth (e.g., a complete set of rules/constraints to identify all errors in the dataset), and simply collects the votes from somewhat reliable, independent sources.

It turns out that we can also describe entity resolution problems in this way by defining $R$ to be a relation whose records are pairs of possibly duplicate records.
Suppose we have a relation $Q$ where some records refer to the same real-world entity.
We can define $R = Q \times Q$ to be the set of all pairs of records from $Q$.
A pair of records $(q_1 \in Q, q_2 \in Q)$ is defined as ``dirty'' if they refer to the same entity, and are clean if otherwise. 
Therefore, $R_{dirty}$ is the set of all duplicated pairs of records (remove commutative and transitive relations to avoid double-counting, e.g., $\{q_1-q_2, q_1-q_4, q_2-q_1, q_2-q_4\} \mapsto \{q_1-q_2, q_1-q_4\}$).
To quantify the number of entity resolution errors in a dataset, we estimate the cardinality of this set. In this paper, we use items, records, and pairs interchangeably.

Some pairs will be obvious matches or non-matches and do not require crowd-sourcing, and in Section \ref{sec:priority}, we describe how to integrate this basic model with algorithmic prioritization.
Furthermore, the above definition only records what are marked as dirty, but does not necessarily correct them. 
While in practice, identifying and correcting are often done together as part of this work we consider them as orthogonal problems.

\subsection{Baselines}
Based on Problem \ref{problem:random}, we describe some existing approaches as our baselines. We consider crowdsourced approaches, in the context where the true number of errors (or a complete set of constraints to identify all violations) is not available.
We categorize these approaches as descriptive or predictive. Descriptive approaches only consider the first $K$ workers (or worker response or task) to clean the dataset, and predictive approaches consider the impact of the future $K'$ (possibly infinite) workers.

\subsubsection{Nominal (descriptive):}

The most basic estimate for the number of errors, is to count the number of records marked as error by at least one worker:
\[
c_{nominal} = nominal(\mathcal{I}) = \sum_{i}^N\mathbbm{1}{[n_i^+>0]}
\]
The number of positive votes on record $i$ (marking it as dirty or an error), $n_i^+$, as well as the number of total votes $n_i$ are at most $K>0$.
We refer to the estimate as $nominal$ estimate. 
However, this estimate is neither forward looking nor tolerant of inconsistency, cognitive biases, and fatigue of workers, and this estimate may be far from the true number $|R_{dirty}|$.

\subsubsection{Voting (descriptive):}
\label{sec:setup:vote}

A  more robust descriptive estimate is the majority consensus:
\[c_{majority} = majority(\mathcal{I})=\sum_{i=1}^N \mathbbm{1}{[n_i^+ - \frac{n_i}{2} > 0]}\]
If the number of workers who marked record $i$ as dirty ($n_i^+$) is more than those who marked it as clean ($n_i^-=n_i-n_i^+$), then we label the record as dirty.
If we assume that workers are better than a random-guesser (i.e., correctly identify dirty or clean records with probability $> 50\%$), then this procedure will converge as we add more workers.
However, for a small number of workers, especially if each worker sees only a small sample of records, even the majority estimate may differ greatly from $|R_{dirty}|$.
Adding more workers may further improve the quality of the dataset, and accordingly, the data quality estimation problem is to estimate the value of adding more workers given $\mathcal{I}$.

\subsubsection{Extrapolation (predictive):}
\label{sec:setup:extrapolation}

Unlike the previous descriptive methods, extrapolation is a simple technique to predict how many more errors exsit in the dataset if more workers were added. 
The core idea of the extrapolation technique is to ``perfectly'' clean a small sample of data and to extrapolate the error rate for the whole data set. 
For example, if a sample of $s= 1\%$ would contain $err_s = 4$ errors, we would assume that the whole data set has $err_{total} =  \frac{1}{s} err_s =  400$ errors or $err_{remaining} =  \frac{1}{s} err_s - err_s = 396$ remaining/undetected errors as we already have identified the $4$ from the sample. 

However, this technique has fundamental limitations: (1) How can one determine that the sample is perfectly clean (i.e., the chicken and egg problem?) and (2) the sample might not be representative. 
More surprisingly, these two problems are even related. 
If the sample is small enough, one can probably put enough effort and resources (e.g., leverage several high-quality experts) to clean the sample.
However, the smaller the size, the higher the chance that the sample is not representative of the original dataset.  
Worse yet, de-duplication requires comparing every item with the every other in the data set (we refer to one comparison between two tuples as an $entity~pair$). 
This makes it even harder to determine a representative subset as we need to retrieve a representative (large enough) sample from an exponential number of combinations ($N * (N-1) / 2$).

\begin{figure}[t]
 \centering
 \includegraphics[width=0.48\columnwidth]{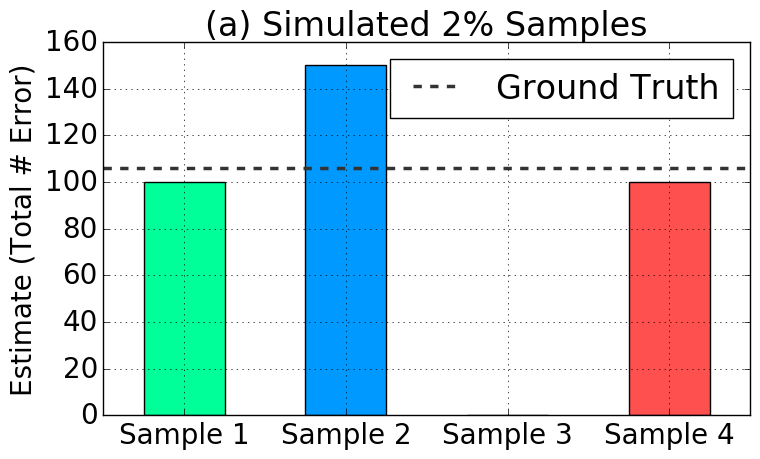}%
 \includegraphics[width=0.48\columnwidth]{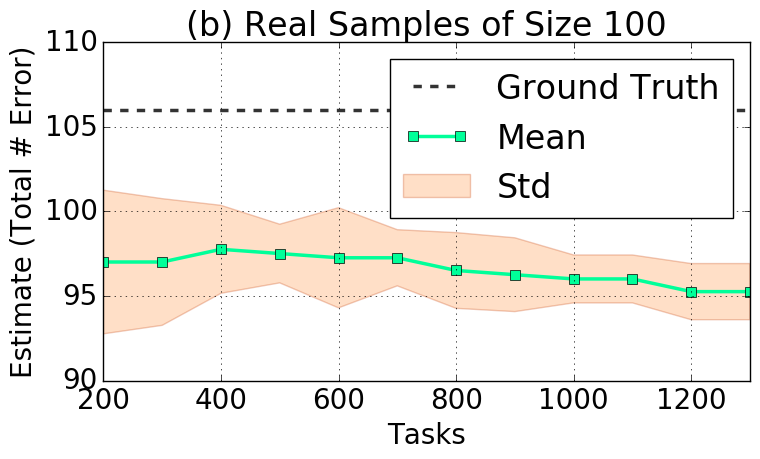}%
  \caption{Estimating the total number of errors: (a) Extrapolation results using four different perfectly cleaned $2\%$ samples; (b) extrapolation results with an increasing effort in cleaning the sample.}
  \label{fig:extrapolation} \vspace{-3mm}
\end{figure}

To illustrate such limitations, we ran an initial experiment using the $restaurant$ data set (also used in \cite{wang2012crowder,gokhale2014corleone}).
The restaurant data set contains 858 records of restaurants with some additional location information:
\vspace{-0.5mm}
\[
\mathbf{Restaurant}(id, name, address, city, category)\vspace{-0.5mm}
\]
Some of the rows of this dataset refer to the same restaurant, e.g., ``Ritz-Carlton Cafe (buckhead)" and ``Cafe  Ritz-Carlton  Buckhead".
Each restaurant was duplicated at most once.
For this data set, we also have the ground truth (the true number of errors), so we are able to quantify the estimation error of different techniques.

First, we run a simulated experiment of directly applying the extrapolation approach to $858 \times 858 = 367653$ entity-pairs (Figure~\ref{fig:extrapolation}(a)).
Of the $367653$ entity pairs, only $106$ of them are duplicate errors, which makes sampling a representative sample (with the representative number of errors) very difficult.
This is related to the known problem of query selectivity in approximate query processing~\cite{DBLP:conf/vldb/HaasNSS95,AgarwalMPMMS13}.
We randomly sampled $2\%$  (about $7300$ pairs) four times and used an ``oracle'' to perfectly clean the data and extrapolated the found errors. 
Figure~\ref{fig:extrapolation}(a) shows how for relatively rare errors, this estimate is highly dependent on the particular sample that is drawn.

This is clearly impractical as cleaning $7300$ pairs perfectly is not trivial, since workers are fallible. Worse yet, even with a perfectly clean sample, the perfect estimate is not guaranteed. Figure~\ref{fig:extrapolation}(b) depicts a more realistic scenario with the previously mentioned 2-stage CrowdER algorithm~\cite{wang2012crowder}. 
We used a normalized edit distance-based similarity and defined the candidate pairs as the ones with similarity between $0.9$ and $0.5$.
Then, we took 4 random samples of size $100$ out of the remaining $1264$ pairs;  we used Amazon Mechanical Turk to have each worker reviewing $10$ to $100$ random pairs from the candidate pairs, taking the majority consensus on items as true labels. Unfortunately for this particular experiment, the (average) estimate moves away from the ground truth as we use more workers to clean the sample. This is because the earlier false positive errors are corrected with more workers.

%% file: species-estimate.tex
\section{Species Estimation Approach}\label{sec:nofp}
The previous section showed that the \naive extrapolation technique does not provide a good estimate. 
The main reason is, that it is very challenging to select a good representative sample (errors are relatively rare) and perfectly clean it. 
Furthermore, perfectly cleaning the sample might be so expensive that it out-weights the benefits of knowing how clean the data set really is.
In this section, we pose the problem as a species estimation problem using a well known estimator called the \textit{Chao92} estimator~\cite{chao1992estimating}.
The ultimate goal is to use species estimation to estimate the diminishing return and through it, the number of remaining errors in the data set. 

\subsection{Overview}

In the basic species estimation problem, we are given $n$ observations drawn with replacement from a population $P$. Suppose, we observe that there are $c$ distinct values in the $n$ samples, the species estimation problem is to estimate $|P|$. 
This problem mirrors the problem of predictive data quality estimates.
Consider a finite population of records (or pairs in the case of entity resolution) a hypothetical infinite pool of workers and for simplicity no false positive errors (i.e., workers might miss an error but not mark a clean record as dirty). 
If we then assume, each worker goes over the whole data set $R$ and marks records in that subset as clean or dirty,  the votes from a worker can be seen as ``discoveries'' of dirty items, i.e, the ``species''.
The species estimation problem is to estimate the total number of distinct erroneous records.

Unfortunately, it is often unrealistic to assume that every worker goes over all records. 
However, without loss of generality we can also present every worker with a random, even differently sized, sub-set of the records. 
As more workers go over $R$, we will still see redundant dirty records marked as dirty by multiple workers as the {\em sample coverage} increases. 
That is, the resulting sample contributed by all the workers remains a sample with replacement.

\subsection{Chao92 Estimator}\label{sec:chao92}

There are several species estimation techniques, and it is well established that no single estimator performs well in all of the settings~\cite{DBLP:conf/vldb/HaasNSS95}. 
In this work, we use the popular \textit{Chao92} estimator, which is defined as:
\begin{equation}
\hat{D}_{noskew} = \frac{c}{\hat{C}}
\label{eqn:chao92noskew1}
\end{equation}

Here, $c$ is the number of unique items (e.g., errors) we observed so far, $C$ the sample coverage, and $D_{noskew}$ our estimate for the total number of unique item. 
Unfortunately, usually the sample coverage is not know. 
To estimate it, the \textit{Chao92} estimator leverages the $f$-statistic, sometimes also referred to as the data fingerprint \cite{valiant2011estimating}. 
The $f$-statistics represent the frequencies of observed data items in the sample, where $f_j$ is the number of data items with exactly $j$ occurrences in the sample. 
For example, $f_1$ refer to all singletons, the errors which were exactly observed ones, whereas the $f_2$, refer to all doubletons, the errors which were exactly observed twice. 
Given the $f$-statistic we can estimate the missing probability mass of all the undetected errors ($f_1/n$); the sample coverage is then estimated as follows: 
\begin{equation}
\hat{C} = 1 - f_1/n
\label{eqn:coverage}
\end{equation}

While it is obvious that $f_1$ refers to the number of times a rare error was observed, it is less clear how $n$ should be defined in this context. 
One might argue that $n$ should refer to the number of total votes by the workers. 
However, we are not interested in votes which declare an item as clean (also recall, that we do not consider any false positives for the moment).
In fact, as we assume that all items are clean and no false positive errors by workers, a negative vote (i.e., clean) is a no-op. 
Thus, $n$ should consist of positive votes $n^+=\sum_{i=1}^{N}{n_i^+}$ only.  This also ensures that $n = n^+ = \sum_{i=1}^{\infty}{f_i}$, yielding to the following estimator: 
\begin{equation}
\hat{D}_{noskew} = \frac{c}{\hat{C}} = \frac{c}{ 1 - f_1/n^+ }
\label{eqn:chao92noskew2}
\end{equation}

We can also explicitly incorporate the skewness of the underlying data as follows:
\begin{equation}
\hat{D}_{Chao92} = \frac{c_{nominal}}{1 - f_1/n^+}  + \frac{f_1\cdot\hat{\gamma}^2}{1 - f_1/n^+}
\label{eqn:chao92}
\end{equation}
where $\gamma$ is \emph{coefficient of variance} and can be estimated as:
\begin{equation}
\label{eqn:cv_est}
\hat{\gamma}^2 = \max \left\{\frac{\frac{c}{\hat{C}} \sum_i{ i(i-1)f_i}}{n^+(n^+-1)} - 1 \, , \,0\right\}
\end{equation}

\subsubsection{Examples and Limitations}

In the following, we show by means examples how false positively negatively impact the prediction accuracy. 

\begin{example} [No False Positives]
Suppose there are $1000$ critical pairs with $100$ duplicates, each task contains $20$ randomly selected pairs. Workers have error detection rate of $0.9$, but make no mistakes (e.g., wrongly mark a clean pair as a duplicate). 
We simulated this scenario and found that after $100$ task, approximately $c_{nominal}=83$ unique errors with $n^+=180$  positive votes and $f_1=30$ errors, which were only identified by a single worker. 
The basic estimate  (without the correction factor $\gamma$) for the number of remaining errors is then:
\[
\hat{D}_{Chao92} - c_{nominal} = \frac{83}{1-30/(180)} - 83 = 16.6,
\]
which is almost a perfect estimate. This is not  surprising as our simulated sampled uniformly with replacement and thus, the Good-Turing estimate will, on average, be exact\footnote{Note, that we do not have a perfect uniform sample, since every task samples the pairs without replacement; however, our results in Section~\ref{sec:exp} show that this is negligible if the number of tasks is large enough. }.
\end{example}

\begin{example} [With False Positives]
Unfortunately, it is unrealistic to assume that there will be no false positives. 
Even worse as errors are rare the number of false positives might be relatively high to the number of actual found errors.
Let's assume $1\%$ chance of false positive error (wrongly classify as dirty). 
Our simulation shows that on average that would add $19$ wrongly marked duplicates, increases the $f_1$ count to $46$ and $n^+=208$. 
With the false positives the estimate changes to:
\[
\hat{D}_{Chao92}  - c_{nominal} = \frac{83 + 19}{1-(46)/(208)} - (83 + 19) \approx 131
\]
Overestimating the number of true errors by  more than $30\%$. 

\end{example}

\subsubsection{The Singleton-error Entanglement}
The reason why the false positives have such a profound impact on the species estimator is two-fold: 
First, they increase the number of unique errors $c$. 
Thus, if we estimate the number of remaining errors, we already start with a higher value. 
Second, and worse yet, the singletons $f_1$ are the best indicator for how many errors are missing, while, at the same time, it is also the best indicator for erroneously classified items (i.e., false positives).
We refer to this problem as the {\bf singleton-error entanglement}.

\subsection{vChao92 Estimator}
\label{sec:voting}

In order to make the estimator more robust, we need a way to mitigate the effect of false positives on the estimate. 
First, we could use $c_{majority}$ instead of $c_{nominal}$ to mitigate the impact on the number of found unique items. 
However, recall that the Chao92 estimate without the correction for the skew is defined as $c_{nominal} / (1 - f_1 / n)$.
That is the estimator is highly sensitive to the singletons $f_1$ qs discussed before. 
One idea to mitigate the effect of false positives, is to  {\bf shift} the frequency statistics $f$ by $s$ and treat $f_{1 +s}$ as $f_1$, etc. 
For instance, with a shift of $s=1$ we would treat doubletons $f_2$ (i.e., the items which were observed twice) as singletons $f_1$ and tripletons $f_3$ as doubletons $f_2$. Shifting $f$ also means that we need to adjust $n^{+,s}=n^+ - \sum_{i=1}^{s}{f_i}$, to ensure the equality of  $n = \sum_{i=1}^{\infty}{f_i}$. 
These statistics are more robust to false positives since they require more workers to mark a record as dirty, but at the cost of some predictive power. Taking the above ideas into account, we derive a new estimator (\textit{vChao92}):
\begin{equation}
\hat{D}_{vChao92} = \frac{c_{majority}}{1 - f_{1+s}/n^{+,s}} + \frac{f_{1+s}\cdot\hat{\gamma}^2}{1 - f_{1+s}/n^{+,s}}
\label{eqn:voting_chao92}
\end{equation}

\textit{vChao92} is more robust to false positives and estimate the size of the remaining errors; however, it converges more slowly than the original sample coverage-based estimators (\textit{Chao92}) due to the singleton-error entanglement. Not only the rare errors $f_1$ are the best indicator for a false positive (i.e., since nobody else marked the same error, it might as well be a mistake), but also they hint on how many more errors are remaining (i.e., if there are many true errors that are hard to identify, then there might be more errors that are undetected because they are relatively more difficult to spot). 
Furthermore, \textit{vChao92} requires to set $s$, which is hard to tune. 
Worse yet, the estimator violates an important (often desired) estimator property: it might not converge to the ground truth. 
In the next section we present another technique, which is parameter free and does not suffer from this problem.

%% file: estimators-fp.tex
\section{Switch Estimation Approach}\label{sec:fp}
In this section, we show how we can estimate a slightly different quantity to avoid the shift parameter $s$ and with better convergence guarantees.

\subsection{Switch Estimation Problem} \label{sec:switch}
Ideally, we want to estimate how many errors are still remaining in a data set. 
In the previous section, we tried to estimate the {\bf total number of errors} based on the initial ``dirty''  data set and the votes (i.e., $\{1,0,\emptyset\}$) from multiple crowd-workers. Now, we ask an alternative question to estimate the {\bf total number of switches}: After we used a cleaning technique (e.g., workers or an automatic algorithm), how many of the initially identified ``clean'' or ``dirty'' items are incorrect? 

This is a different problem since we are no longer trying to just estimate  the total number of errors (``dirty'' items). Instead, we estimate the wrongly marked items: {\bf At any given time, how many wrongly marked items (false positives and false negatives) does the data set still contain?}  Assuming that workers are, on average, better than a random guesser, the majority consensus on each item will eventually be correct with a sufficient number of workers and their responses. 
This observation allows us to rephrase the original problem in terms of the consensus: {\bf At any given time, how many of the majority vote decisions for any given record do we expect to switch?} 

\begin{problem}[Switch Estimation]\sloppy 
We estimate the number of expected switches for the current majority consensus vector $\mathcal{V} \in \{0,1\}^N$ to reach the ground truth vector $\mathcal{E} \in \{0,1\}^N$.
The data quality estimation problem is to estimate $|\{(v_i,e_i):v_i \ne e_i, v_i \in \mathcal{V}, e_i \in \mathcal{E}\}|$ given $\mathcal{V}$, but not $\mathcal{E}$.
\label{problem:switch}
\end{problem}

Switches act as a proxy to actual errors and, in many cases, might actually be more informative.
However, since a record can switch from clean to dirty and then again from dirty to clean, it is not the same as the amount of dirty records or remaining errors in the data set. 
We define the number of switches in $\mathcal{I}$ as follows:
\begin{equation}\label{eq:switch}
   switch(\mathcal{I}) = \sum_{i=1}^{N}\left( \sum_{j=2}^{K} {\underbrace{\mathbbm{1}[n_{i,1:j}^+ = n_{i,1:j}^-]}_{i} }+ \underbrace{\mathbbm{1}[n_{i,1}^{+} = 1 ]}_{ii}\right)
\end{equation}

Here, $n_{i,1:j}^+$ (or $n_{i,1:j}^+$) denotes the number of positive (or negative) votes on $i$ among workers $1$ through $j$.
Assuming a switch happens every time, there will be a tie in the votes (part i of equation~\ref{eq:switch}). Yet special attention has to be given to the beginning (i.e., the first vote). 
Here, we assume that all data items in the beginning are clean, and that if the first vote for a record is positive (e.g., marks it as dirty), the record switches (part ii of equation~\ref{eq:switch}). 
However, it is easy to extend the switch counting definition to consider various policies (e.g., tie-breaking). 

\subsection{Remaining Switch Estimation}
\label{sec:switch:remaining}
While problem re-formulation is promising (Problem~\ref{problem:switch}: it better reflects the number of remaining errors in the presence of human errors), it poses one significant challenge to our species estimation techniques:
How should we define the $f$-statistics in the problem context? 
Surprisingly, there exists a simple and sound solution. 
Whenever a new switch occurs on item $i$, it is a $singleton$.
If we receive an additional vote on $i$ without flipping the consensus, we say the {\em switch is rediscovered}, and it changes to a $doubleton$.
If another vote, yet again, does not flip the consensus, the same switch gets rediscovered again (e.g., changes it from a $doubleton$ to a $tripleton$). 
On the other hand, if the consensus on $i$ flips from, say, clean to dirty and then clean again, we do not rediscover the same switch, but a new one.

It naturally follows that we define $n$ as the sum of the frequencies of the switch frequencies, $n = \sum_{i=1}^{\infty}{f_i}$. While this preserves the relationship between $f_1$ and $n$ in the original species estimation technique, we found that this definition has a tendency to overestimate the number of switches. 
The reason is that, with every new switch, it is implicitly assumed that the sampling for the item resets and starts sampling for a new switch (although it is still for the same item) from scratch again. 
Therefore, we use a small modification and simply count all votes as $n$. 

Finally,  we need to, again, pay special attention to the first votes.
Ideally, we would first get at least three votes for every item before doing any estimation. 
However, in practice we often want to start estimating based on the default labels (e.g., all records are assumed to be clean) before receiving a lot of votes for every record.  
The question is how votes that stays with the default labels (i.e., votes before the first switch) should be counted. 
Again, the solution is rather simple: The first $k$ votes, which do not cause the default label to switch, do not re-discover a switch, and thus, they are no-ops (i.e., do not contribute to $f$-statistics as well as to $n$). 
Hence, we need to adjust $n$ based on this new assumption and subtract all no-ops: 
\[
n^{switch}=n- \sum_{i=1}^N{(\argmin_{j\in [1,K]}\{n_{i,1:j}^+\geq n_{i,1:j}^- \}-1) }
\]
The last term of the equation removes all the no-ops before the first switch on each items (i.e., votes prior to the first switch do not contribute to the $f$-statistics, so they should to $n$).

We can now estimate the total number of switches as $K \rightarrow \infty$, using the same $Chao92$ estimation technique:
\begin{equation}
\hat{D}_{Switch} = \frac{c_{switch}}{1 - f'_{1}/n^{switch}} + \frac{f'_{1}\cdot\hat{\gamma}^2}{1 - f'_{1}/n^{switch}}
\label{eqn:switch_chao92}
\end{equation}
where $c_{switch}$ is the number of records with any consensus flips, i.e., switches:
\vspace{-2mm}
\[
c_{switch}=\sum_{i=1}^N{\mathbbm{1}[switch(\mathcal{I}_{i,1:K})>0]}.
\]
Using this estimator, then the expected number of switches needed for the current majority consensus to reach the ground truth is
\[
\xi = \hat{D}_{Switch} - switch(\mathcal{I}).
\]

This estimator has several desired properties. First, it is parameter-free, and there is no $s$ to adjust. 
Furthermore, the estimator is guaranteed to converge to the ground truth. 
This is because of our main assumption, where the majority vote will eventually become the correct labels (i.e., workers are better than a random guesser). 
The more workers whow confirm to the majority consensus, the fewer switches to be expected, and the estimator will predict the lower number of remaining switches (consider the definition of our  $f'$-statistic). 
Lastly, as the estimator is more robust against false positives, it becomes less likely that, as the number of votes per item increases, a false positive wwill flip the consensus.

\subsection{Switch-Based Total Error Estimation}
\label{sec:fp:total}

Interestingly, although we changed the estimation problem from trying to estimate the total number of errors (Problem~\ref{problem:random}) to how many switches we expect from the current majority vote (Problem~\ref{problem:switch}), we can still use the switch estimation to derive an estimate for the total number of errors (Problem~\ref{problem:random}).  
 The idea is that we adjust $majority$ by the number of positive and negative switch estimates:
\[
majority(\mathcal{I}) + \xi^+ - \xi^-
\]
where positive switch $\xi^+$ is defined as switches from the ``clean'' label to the ``dirty'' label and negative switch $\xi^-$ as switches from ``dirty'' to ``clean.''
This only requires that we count $c_{switch}$ and $f1'$ based on positive (or negative) switches only and estimate $\xi^+$ (or $\xi^-$) using the switch estimator (Equation~\ref{eqn:switch_chao92}) . 

This makes sense, except that a separate estimation of positive and negative switches can cause either the positive or negative switch estimation to fail due to a lack of observed switches. To mitigate this problem, we make a key observation that $majority$ tends to improve monotonically with more task responses. Therefore, if we detect an increasing trend in $majority$, then it means that we have more positive switches and would have even more due to the monotonicity. If this is the case, then we focus on the positive switch estimates to adjust $majority$ total error estimate as:
\[
majority(\mathcal{I}) + \xi^+
\]
Similarly, if we observe a decreasing $majority$ trend, then we can focus on the negative switch estimate:
\[
majority(\mathcal{I}) - \xi^-
\]
Our final total error estimation technique makes this decision dynamically to improve the estimates.

%% file: priotization.tex
\section{Estimation With Prioritization}\label{sec:priority}
A common design paradigm for data cleaning methods is ``propose-verify''. In a first step, a heuristic identifies candidate errors and proposes repairs. Then, a human verifies whether the proposed repair should be applied. This can further be extended where a human only verifies ambiguous or difficult repairs.
In this section, we discuss the case where the data are not sampled uniformly.

\subsection{Prioritization and Estimation}
For some types of error, randomly sampling records to show first to the crowd will be very inefficient. Consider a crowdsourced Entity Resolution scenario, where the crowd workers are employed to verify matching pairs of records. Out of $N$ total records, suppose $k$ have exactly one duplicate in the dataset. This means that out of $\frac{N(N-1)}{2}$ pairs only $k$ are duplicate pairs--meaning that even though the probability of sampling a record that is duplicated is $\frac{k}{N}$, it is roughly $\frac{k}{N^2}$ to sample a duplicated pair. It would be infeasible to show workers a sufficiently rich random sample for meaningful estimation in large datasets.

It is often the case the crowd sourced data cleaning is run in conjunction with algorithmic techniques.
For example, in Entity Resolution, we may merge records that are sufficiently similar automatically, and reserve the ambiguous pairs for the crowd.
To formalize, there exists a function $H: R \mapsto \mathbb{R}_+$ that is a measure of confidence that a record is erroneous.
We assume that we are given a set of ambiguous records selected by the heuristic $R_H = \{\forall r \in R: \alpha \le H(r)  \le \beta\}$, and this section describes how to utilize $R_H$ in our estimates.

\subsection{Simple: Perfect Heuristic}
First, we consider the case where the heuristic $H$ is perfect, that is, $\{r \in R: H(r) > \beta\} \cap  R_{dirty}^c = \emptyset$ and $\{r \in R : H(r) < \alpha\} \cap R_{dirty} = \emptyset$. 
Note that the number of true errors in $R_H$ is less than or equal to $|R_{dirty}|$, i.e., $R_H$ does not contain any obvious cases, $R_H^c = \{\forall r \in R: H(r) < \alpha ~or~ H(r) > \beta\}$. 
It turns out that this is the straight-forward case, and the problem is essentially as same as the original estimation problems  (Problem~\ref{problem:random}, \ref{problem:switch}).

In the case of perfect heuristic, we randomly select a sample of $p$ pairs from $R_H$ and assign a number of workers to the sample; overall with $K$ workers, we have $n= p \cdot K$ pairs/records marked by workers. 
Now, the target estimate is: $\hat{D}(R_H)=|R_{dirty} - R_H^c|$ instead of $|R_{dirty}|$. This result is easy to interpret as the perfect heuristic guarantees that $\{r \in R: H(r) > \beta\} \cap  R_{dirty}^c = \emptyset$ and $\{r \in R : H(r) < \alpha\} \cap R_{dirty} = \emptyset$:
\begin{equation}\label{eqn:perfect_heuristic}
|R_{dirty}| = \hat{D}(R_H) + |\{r \in R: H(r) > \beta\}|
\end{equation}
$\hat{D}(R_H)$ can be either $\hat{D}_{vChao92}$ or $\hat{D}_{GT}$ on $R_H$. Note that, depending on the qualities of workers, we still have false positive and false negative errors; the workers might make more mistakes as less obvious are asked. The obvious cases are efficiently classified by algorithm techniques.

\subsection{Harder: Imperfect Heuristic}\label{sec:imp_heuristic}

Next, we consider the harder case where the heuristic is imperfect: $\{r \in R: H(r) > \beta\} \cap R_{dirty}^c \neq \emptyset$ and $\{r \in R : H(r) < \alpha\} \cap R_{dirty} \neq \emptyset$. In this case, we would not only have false positives and false negatives on $R_H$ due to the workers, but also have false negatives in $R_H^c$ missed and false positives in $R_H^c$ incorrectly identified by the heuristic itself. 

In particular, we cannot use the simple approach (Equation~\ref{eqn:perfect_heuristic}) as with perfect heuristic, since $|\{r \in R: H(r) > \beta\}|$ might contain false positives by algorithm techniques. 
For example, if the heuristic's upper threshold $\alpha$ is too loose (with many false positives), we may not see any new errors in $R_H$.
Moreover, we also need to include false negatives in  $\{r \in R : H(r) > \beta\}$ to get the ground truth $|R_{dirty}|$. 

To address this problem, we employ randomization.
Workers see records from $R_H$ with some probability $1-\epsilon$ and see records from $R_H^c$ with probability $\epsilon$.
This allows us to estimate remaining errors in $R_H$ and $R_H^c$ using the proposed techniques:
\begin{equation}\label{eqn:imperfect_heuristic}
|R_{dirty}| = \hat{D}(R)
\end{equation}
Even though we are estimating over $R$, we still ask workers to look at mainly examples from $R_H$. The idea is that $R_H^c$ still contains a few true errors compared to $R_H$, and we need a fewer crowd answers to perform accurate species estimation.
$\epsilon$ can be thought of as a measure of ``trust" in the heuristic.
As $\epsilon$ goes to zero, the model limits to the perfect case.
On the other hand, as $\epsilon$ goes to $\frac{|R_H|}{|R|}$ the model limits to the random sampling case\footnote{$\frac{|R_H|}{|R|} < \epsilon < 1$ implies that the Heuristic is negatively correlated with errors}.
Thus, $\epsilon$ defines a family of tunable data quality estimators that can leverage user-specified heuristics.
In our experiments, we found that $\epsilon = 0.1$ is a good value to use.

%% file: experiment.tex
\section{Experiments}
\label{sec:exp}

We evaluate our estimation technique using real world crowd-sourced data sets as well as synthetic ones. We designed our experiments to address the following questions:
\begin{itemize}
\setlength\itemsep{0em}
\item Do techniques perform on real-world data sets? 
\item What is the sensitivity of our estimators to false positive and false negative errors?
\item Is reformulated switch estimation more robust to false positive errors?
\item How useful is prioritization in error/switch estimation?
\end{itemize}

\subsection{Real-World Data Sets}\label{sec:real_exp}
We used Amazon Mechanical Turks (AMT) to crowd-source entity resolution tasks. Each worker receives a task consisting of a number of candidate records sampled from a data set $R$; the worker identifies which of the records are duplicates (label as {\em dirty}), for $\$0.03$ per task (e.g., $\$3.00$ per 100 tasks). Each task contains 10 records and a worker may take on more than a single task. In an effort to reduce noise in the collected responses, we designed a qualification test to exclude workers who are not familiar with the contexts of the data sets. 
When presenting results, we also randomly permute the workers and average the results over $r=10$ such permutations.
This averages out any noise in the particular order of tasks seen by workers.
We consider the following approaches: \textbf{V-CHAO}, the voting and shift-based estimator with shift $s=1$ (Section~\ref{sec:voting}); \textbf{SWITCH}, a total error estimate derived from the switch estimator (Section~\ref{sec:switch:remaining}); \textbf{VOTING}, the current majority vote over all items (Section~\ref{sec:setup:vote}); and \textbf{Ground Truth}, the true number of errors (or switches needed). We also show \textbf{EXTRAPOL} (Section~\ref{sec:setup:extrapolation}) as a range of +/- 1-std around the mean of extrapolation results based on a perfectly clean $5\%$ sample (assuming it was provided by an oracle). 
Lastly, we also show what we call Sample Clean Minimum (\textbf{SCM}), or the minimum number of tasks needed to clean the sample with a fixed number of workers per record: 
\[
\frac{3~worker \times S~records}{p~records/task \times 1~task/worker},
\]
where we assume three workers to each record in the sample of size $S << N$, and each task contains $p$ records and is assigned to a single (independent) worker.
The point is to illustrate that the proposed DQM is comparable in the cost (the number of tasks) to actually cleaning the sample with a fixed number of workers, without the added redundancy of worker assignment. Of course, a perfectly clean sample does not provide reliable estimates like the proposed technique.

\begin{figure*}[t!]
 \centering
 \includegraphics[width=0.33\linewidth]{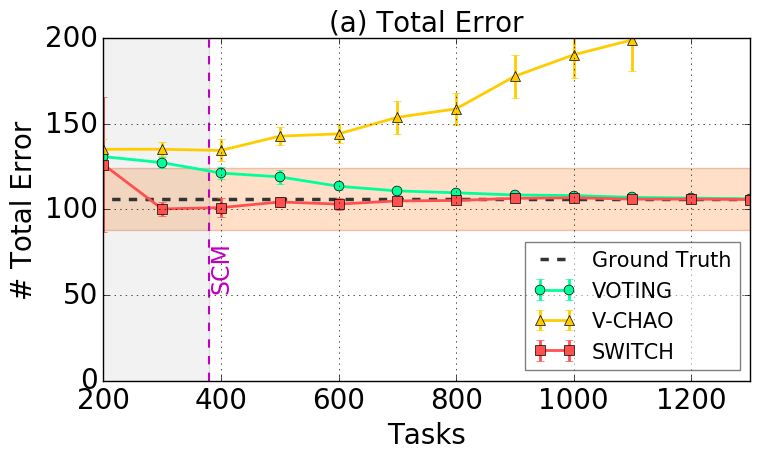}
 \includegraphics[width=0.33\linewidth]{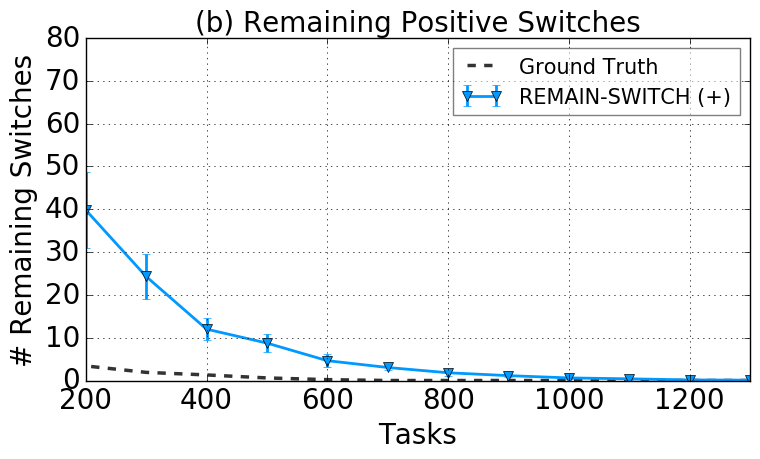}
  \includegraphics[width=0.33\linewidth]{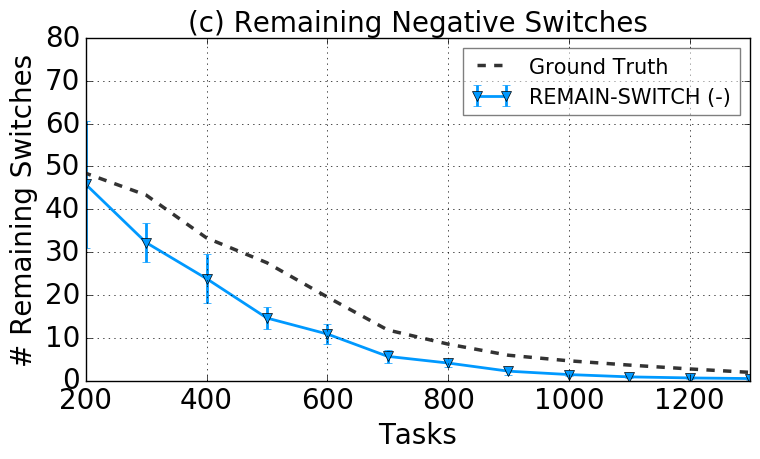}
  \caption{Total error estimation and positive and negative switch estimation results on {\bf Restaurant Data Set}: there are more false positive errors, so \textit{VOTING} monotonically decreases; hence,  \textit{SWITCH} provides the most accurate total error estimates based on the negative swith estimation.  
 \label{real:restaurant}}\vspace{-3mm}
\end{figure*}
\begin{figure*}[t!]
 \centering
 \includegraphics[width=0.33\linewidth]{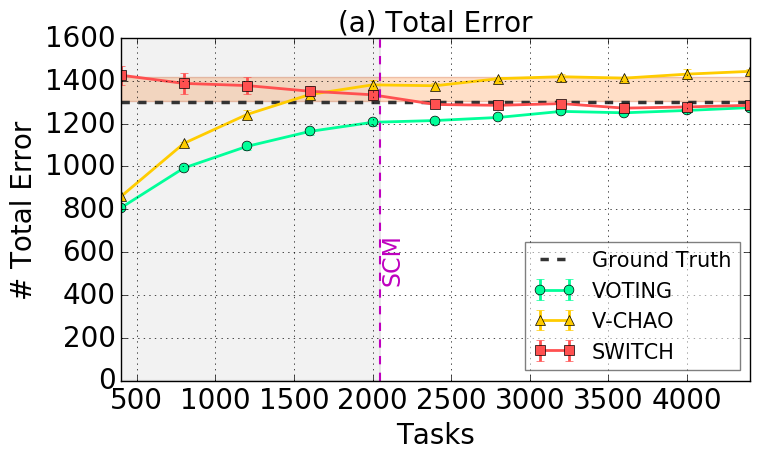}
   \includegraphics[width=0.33\linewidth]{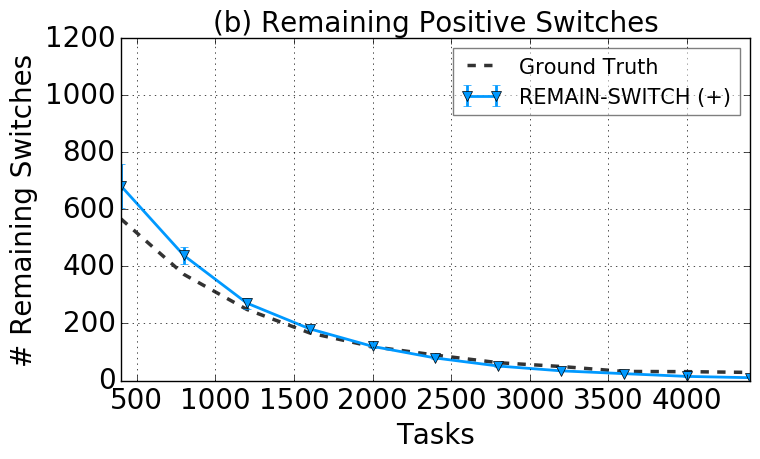}
   \includegraphics[width=0.33\linewidth]{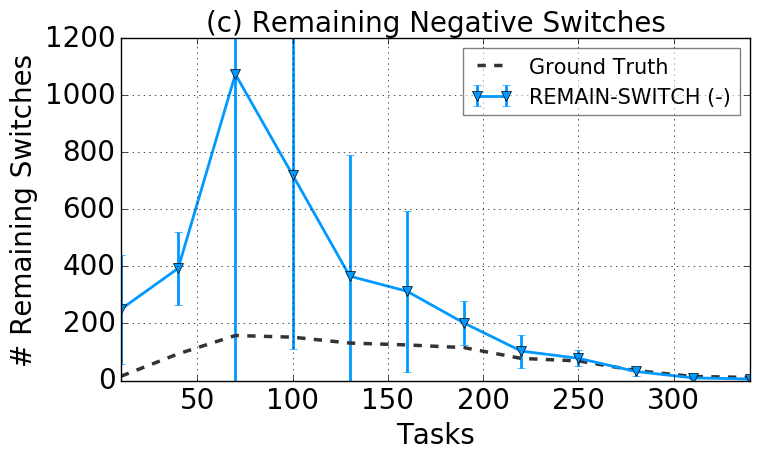}
  \caption{{\bf Product Data Set} contains more false negative errors and \textit{SWITCH} uses the more accurate positive switch estimation to yield the most accurate total error estimates.} \label{real:product} \vspace{-3mm}
\end{figure*}

\subsubsection{Restaurant Data set}
We experimented on the restaurant dataset described earlier in the paper.
The heuristic that we used to filter the obvious matches and nom-matching pairs was a normalized edit distance-based similarity $> 0.9$ and $< 0.5$. 
The remaining candidate pairs among $858\times 858$ total pairs are $1264$ pairs, with $12$ pairs being true duplicates. Each task consists of ten random pairs from the $1264$ critical pairs, which is given to a worker.
Figure~\ref{real:restaurant} shows how different estimation techniques work on the data set.

Given samples from the candidate pairs, the workers make a lot of false positive errors; the species estimation technique \textit{V\textnormal{-}CHAO} fails because the $f_1$ and $f_2$ counts are over-inflated. However, the switch estimation is more robust to false positives; thus \textit{SWITCH} traces the ground truth very nicely and with a fewer number of tasks than the current majority vote (\textit{VOTING}).
So, even before all the pairs have been marked as clean or dirty (duplicate), we can already make a good prediction about the number of errors (duplicates). 
\textit{EXTRAPOL} and its one standard deviation band illustrates that  extrapolation based on a small perfectly clean sample does not provide reliable estimates due to the high variability. Unless you take many perfectly clean samples to average out the estimates, this is expensive and impractical; other estimates (e.g., \textit{SWITCH}) are based on just a single sample, with some false positive and false negative labels. 
Lastly, the species estimation techniques require added redundancy due to the random assignment of workers with overlaps. But because \textit{SWITCH} can estimate the number of remaining/total errors correctly, even before covering all the pairs within the sample, the number of tasks needed for \textit{SWITCH} to provide good estimates is comparable to \textit{SCM}. This is encouraging, especially because cleaning a sample perfectly and efficiently without redundancy would not guarantee good estimates.

Figure~\ref{real:restaurant}(b) and (c) show the number of positive and negative switches over time (i.e., tasks).
The ground truth in switch estimation is defined as the number of switches needed (split in positive and negative) for the current consensus on all candidate pairs to reach the true labels. 
As it can be seen, there are more negative switches than positive switches left. 
This implies that we have more false positives (i.e., wrongly marked duplicate pairs) than false negatives.
Another indication of the number of increased false positives is that \textit{VOTING} in Figure~\ref{real:restaurant}(a) decreases over time. 
As we described in Section~\ref{sec:fp:total}, \textit{SWITCH} uses the number of estimated negative or positive switches to adjust the current majority consensus result.
For the restaurant data set, we observe that the \textit{SWITCH} estimates starts to exclusively use the estimate for the negative switches early on as the number of records marked as dirty is monotonically decreases. 
Figure~\ref{real:restaurant}(b) and (c) show that this is also the right choice. 
The estimation mechanism of \textit{SWITCH} exploits the fact that the dataset error rate monotonically improves (aka diminishing utility of error cleaning) over the increasing cleaning efforts; \textit{SWITCH} uses either positive or negative switch estimates to correct \textit{VOTING}. This results in a couple of nice properties for \textit{SWITCH}. First, \textit{SWITCH} makes the correction based on more reliable switch estimates. Tasks often consist of a single type of switch, which means that only positive or negative switch estimates are more reliable. Second, the monotonicity allows \textit{SWITCH} to be at least as good as the baseline, since it uses positive switch (adding errors) only if \textit{VOTING} is increasing and negative switch (subtracting errors) if \textit{VOTING} is decreasing. This also means that, \textit{SWITCH} can converge with the help of only one-sided switch estimates (negative switch estimation in this case).

\begin{figure*}[t!]
 \centering
 \includegraphics[width=0.33\linewidth]{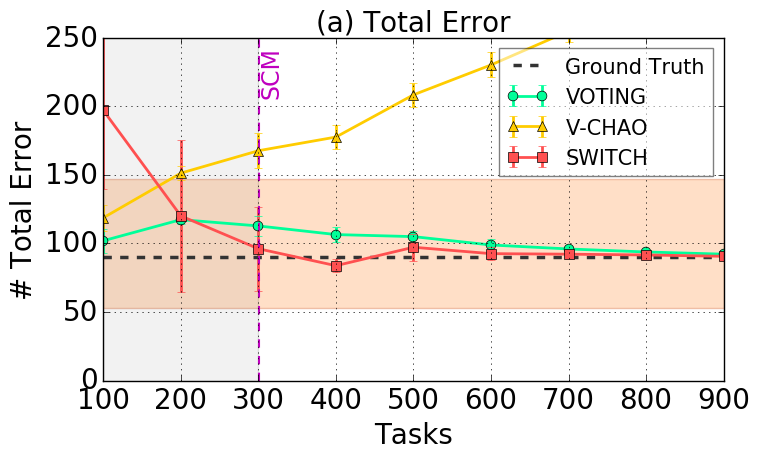}
   \includegraphics[width=0.33\linewidth]{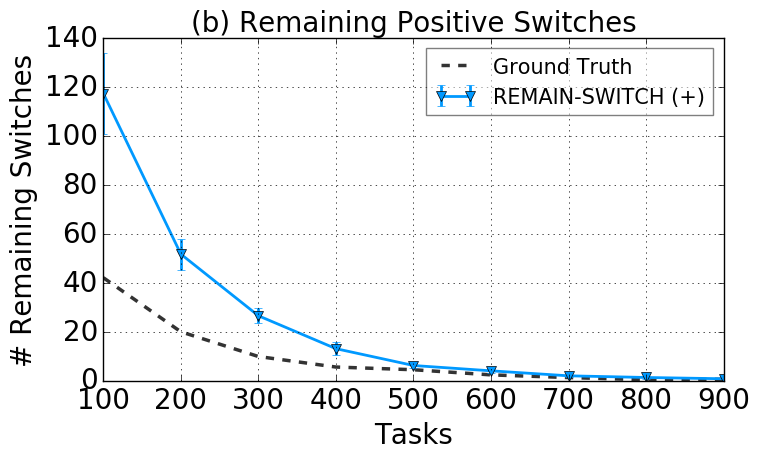}
   \includegraphics[width=0.33\linewidth]{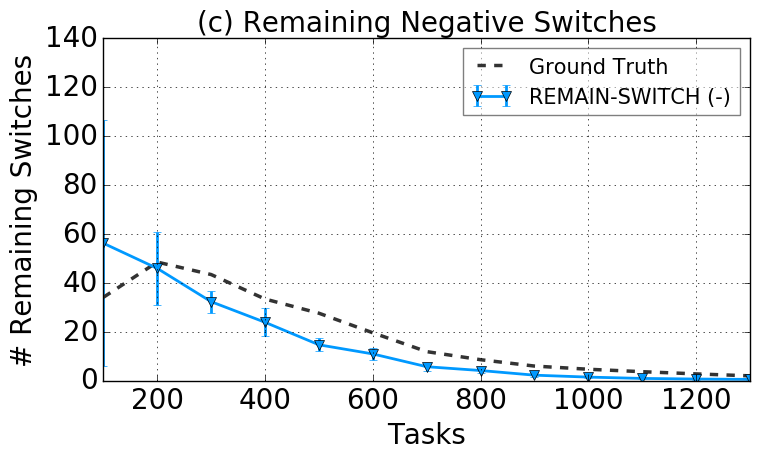}
  \caption{{\bf Address Data Set} contains both false positve and negative errors in fair amounts. \textit{SWITCH} uses positive switch estimates initially and overestimates further; however, once the workers slowly begin to correct false positive errors, \textit{SWITCH} quickly starts to converge to the ground truth leveraging the accurate negative switch estimates.} \label{real:address} \vspace{-3mm}
\end{figure*}

\subsubsection{Product Data Set}
The product data set~\cite{wang2012crowder,gokhale2014corleone} contains 2336 records by Amazon and 1363 records by Google in the form:
\[
\mathbf{Product}(retailer, id, name1, name2, vendor, price) \vspace{-1mm}
\]
Each row belongs to either Amazon or Google ($retailer=\{Amazon, Google\}$), and the size of possible pairs is $1363\times 2336$. 
Each product has at most one other matched product.
The heuristic we used to filter the obvious matches and non-matching pairs was a normalized edit distance-based similarity $> 0.7$ and $< 0.4$. 
Thus, the prioritized candidate pairs contain $13022$ pairs, of which $607$ pairs are true duplicates. Because the matching task is more difficult than that of the restaurant data set, we observed more mistakes from workers. 

Unlike the Restaurant dataset, the Product dataset contains more false negative errors; hence, \textit{VOTING} exhibits an increasing trend, and we observe more positive switches. Therefore, 
\textit{SWITCH} leverages remaining positive switch estimates. 
Figure~\ref{real:product} shows \textit{SWITCH} outperforms all other estimators. 
Again, compared to the current majority votes, it provides a good estimate for the total number of errors early on. 
It is interesting to see that \textit{V\textnormal{-}CHAO} performs reasonably well in the early stage ($< 1200$ tasks); however, its error sharply increases as we add more tasks. This can be attributed to the fact that we have a few difficult pairs on which more than just a single worker make mistakes. In this case, a fixed shift $s=1$ cannot prevent \textit{V\textnormal{-}CHAO} from overestimating. Coming up with the optimal shift parameter $s$ a priori is a challenge, especially as it will vary across different data sets and worker responses.

Figure~\ref{real:product}(b) and (c) show, as before, the separate estimates for the number of remaining positive and negative switches. This time, the positive switch estimation is more accurate; the negative switch estimation over-estimates and has rather large error bars.
This indicates a low number of existing negative switches observed to perform reliable estimations. 
Therefore, as \textit{VOTING} increases monotonically, \textit{SWITCH} uses only the remaining positive switch estimates, and the heuristic yields a good total/remaining number of errors estimates as seen in Figure~\ref{real:product}(a). \textit{EXTRAPOL}, visualized through the shaded area in Figure~\ref{real:product}(a), highly varies in the quality. The reason is that, again, the relatively small random sample may or may not be representative of the true error distribution. 
Furthermore, it should be noted that perfectly cleaning even a small $5\%$ sample ($13022\cdot 0.05 \approx 651$ pairs) is often already impractical (i.e., \emph{chicken-and-egg problem}); how do we know after cleaning if the sample is perfectly clean?

\subsubsection{Address Data Set}\label{sec:mal_addr}
The address data set contains 1000 registered home addresses in Portland, OR, USA, which confirms to the following format in the given order: 
\[
<number~street~unit,~city,~state,~zip>
\]
The house number ($unit$) is optional, and the task is to identify any malformed address entries. The data set contains 90 errors (misformatted records). Since the task does not require pairwise comparison and the number of candidate entries is reasonable, we did not impose any prioritization rules. 

This experiment is interesting for a couple of reasons. First, it deals with a different type of error (i.e., misformatted data entry). Second, the data set contains both false positives and false negatives in fair amounts, so \textit{VOTING} does not improve much initially, for up to 300 tasks. In response, \textit{SWITCH} uses positive switch estimates initially and overestimates further. However, once the workers slowly begin to correct false positive errors, \textit{SWITCH} quickly starts to converge to the ground truth.
Figure~\ref{real:address} illustrates this, along with both positive and negative remaining switch estimation results.

As a result, \textit{SWITCH} estimates first  based on the number of remaining positive switches and then on the remaining negative switches later. But it has a tendency to overestimate early on due to the large number of initial false positives. 
However, after gathering enough tasks (\textit{SCM}), the \textit{SWITCH} estimates converge to the true number of errors in the dataset.

\subsection{Simulation Study}\label{sec:simulation}
Next, we evaluate the different estimation techniques in a simulation based on the $Restaurant$ data set. 
For all simulations, we used a subset with 1000 candidate pairs, among which 100 pairs are true duplicates. 
We randomly generated tasks and worker responses with the desired worker accuracy (precision) and task sampling rate (coverage); there are three types of workers: namely, a false positive errors only-worker, false negative errors only-worker, or worker that make both types of errors. When a direct comparison of the original estimates is not appropriate (e.g., \textit{Chao92} heavily overestimates with false positives), we used a scaled error metric, $SRMSE=\frac{1}{D}\sqrt{\frac{1}{r}\sum_{r}{(\hat{D}-D)^2 }}$, to compare widely varying total error estimates of different techniques. Again, we permute the simulated data to repeat the experiments $r=10$ times.

\subsubsection{Sensitivity of Total Error Estimation}
One of the central claims of this paper is that species estimation is not robust to false positive errors. To illustrate this point, we first explore exactly where the trade-off point would be between the two types of errors.

\begin{figure}[h!]
 \centering
 \includegraphics[width=0.5\columnwidth]{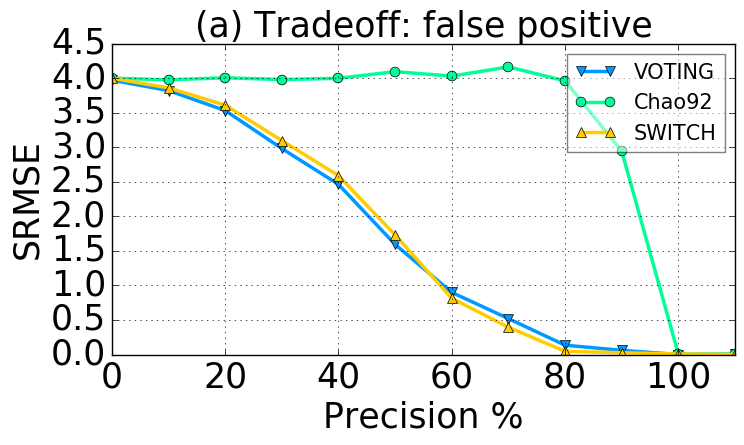}
  \includegraphics[width=0.5\columnwidth]{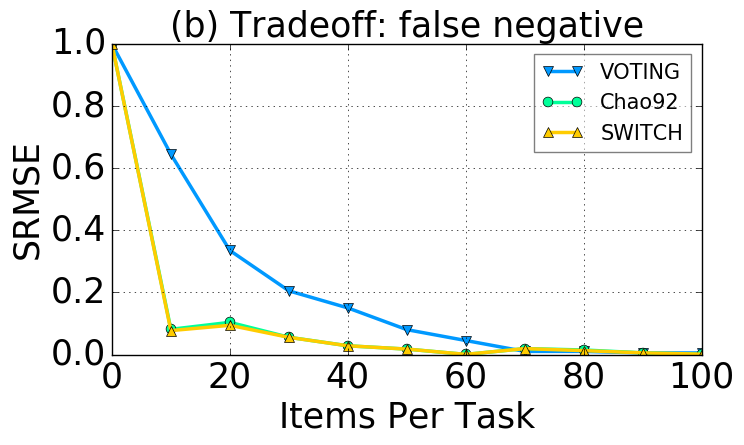}
  \caption{For a fixed number of tasks, we measure the scaled errors of the estimates as a function of (a) worker quality (precision) and (b)  number of items per task (coverage). \label{sim:exp1}} \vspace{-2mm}
\end{figure}

\begin{figure*}[t!]
 \centering
 \includegraphics[width=\textwidth]{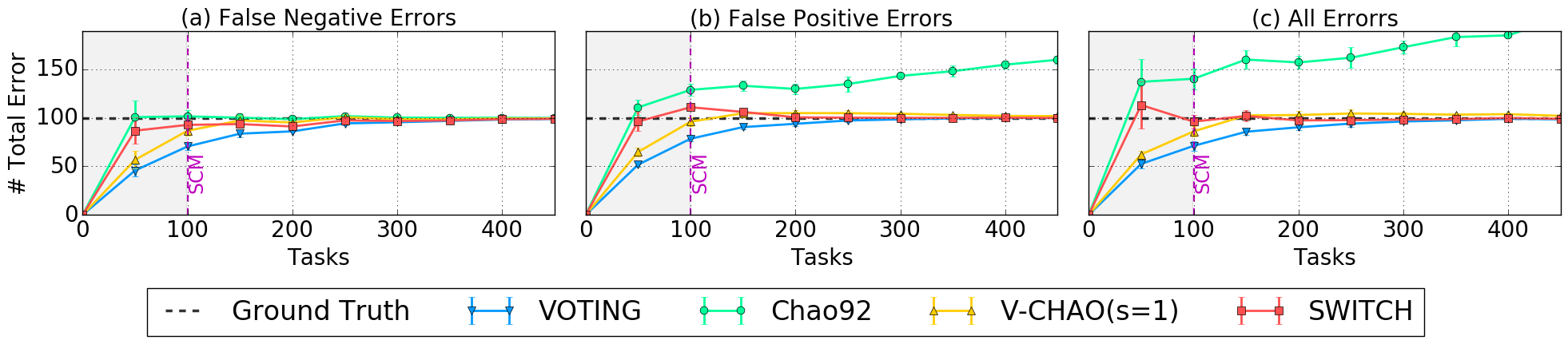}
  \caption{Total error estimates using the simulated datasets; \textit{SWITCH} is the most robust estimator against all error types.  \label{sim:exp2}} \vspace{-5mm}
\end{figure*}

In Figure \ref{sim:exp1}(a), we vary precision (i.e., the failure rate of workers) for the given 50 tasks, each containing 15 items; in (b), we vary the number of items per task from 0 to 100.
What can be seen in Figure~\ref{sim:exp1}(a) is that the \textit{Chao92} is very sensitive to the number of false positives whereas \textit{SWITCH}  follows \textit{VOTING} more closely.
Furthermore, with a higher precision, \textit{SWITCH} does slightly better than \textit{VOTING} (e.g., it has the correct predictive power). But once workers becomes more fallible ($precision<50\%$), \textit{VOTING} becomes slightly better because there is really nothing that the \textit{SWITCH} estimator can do (i.e., our main assumption that the majority consensus eventually converges to the ground truth is violated). 

On the other hand, Figure~\ref{sim:exp1}(b) shows that if there are no false positives, \textit{Chao92} does very well. 
This is not surprising as the estimation technique is forward looking (i.e., robust to false negatives). 
However, only \textit{SWITCH}  is capable of dealing with both false positives and false negatives.

\subsubsection{Switch Estimation Is More Robust}
We now study  how the switch estimator behaves in comparison to the original species estimatiors (\textit{Chao92},  \textit{V\textnormal{-}CHAO}).
We consider three scenarios: 1) only with a false negative rate of $10\%$  (e.g., a $10\%$  chance that a worker overlooks a true error), 2) only with a false positive rate of $1\%$, and 3) with both false negative and false positive errors at rates of $10\%$ and $1\%$, respectively. 
Each task contains $15$ items. 

Figure~\ref{sim:exp2}(a) shows, without any false positives, that \textit{Chao92} performs the best, and all other estimators converge more slowly. 
However, \textit{SWITCH} still converges much faster than  \textit{V\textnormal{-}CHAO}. The picture changes quite a bit in the case of false positive errors (b).  
\textit{Chao92} now strongly overestimates, whereas  \textit{V\textnormal{-}CHAO} and \textit{SWITCH} provide accurate estimates. Here, \textit{V\textnormal{-}CHAO} actually performs surprisingly well. 
The reason is that, in our simulation, the error is evenly distributed across the items, making the (f-statistics) shifting more effective; an assumption that, unfortunately, rarely holds in practice as our real-world datasets experiments have demonstrated. Finally, in the case of both type of errors (c), \textit{SWITCH} again performs well while  \textit{Chao92} again strongly overestimates and  \textit{V\textnormal{-}CHAO} takes longer to converge and also slightly overestimates. 

\subsubsection{Prioritization}
\begin{figure}[h!]
 \centering
 \includegraphics[width=0.495\columnwidth]{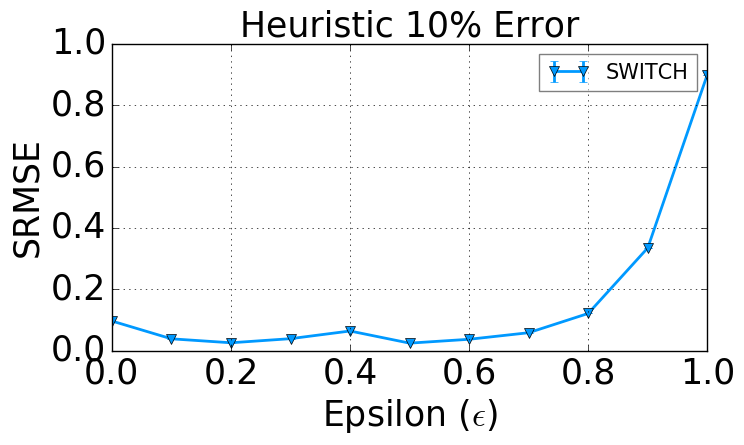}
  \includegraphics[width=0.495\columnwidth]{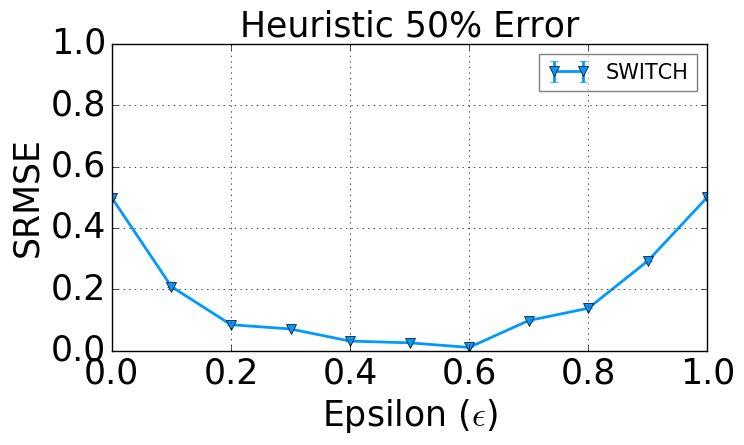}
  \caption{For a fixed error rate and 50 tasks, we measure the accuracy of the switch estimate as a function of the quality of the heuristic ( $\epsilon$). \label{sim:exp5}} \vspace{-2mm}
\end{figure}

In our experiments on real data, we showed how our estimation techniques can be coupled with heuristics to prioritize what to show to the workers.
However, sometimes, the heuristics may be imperfect (Section~\ref{sec:imp_heuristic}).
To address this issue, we employ randomization.
Workers see records from $R_H$, with some probability $1-\epsilon$, and  records from $R_H^c$ with probability $\epsilon$.
We measure the sensitivity of our best approach (\textit{SWITCH}) to the choice of $\epsilon$ in Figure \ref{sim:exp5}.
For a fixed error rate and 50 tasks, we measure the accuracy of the estimator as a function of $\epsilon$, the quality of the heuristic.
We have a heuristic that has a 10\% error rate  and one that has a 50\% error rate.
When the heuristic is mostly accurate (10\% heuristic error) small settings of $\epsilon$ suffice, and it is better to set $\epsilon$ lower.
On the other hand, when the heuristic is very inaccurate (50\% heuristic error), problems can arise.
In our real experiments, we found that standard similarity metrics work very well as heuristics for de-duplication tasks.

\subsection{Trust In The Results}
The proposed data quality metric (DQM) is the first step to measure data quality, in terms of any undetected remaining errors, without the ground truth (e.g., prior set of rules or constrainsts to define the perfect state of the perfect dataset). This is an important problem in data exploration as data scientists often lack the complete domain knowledge or such constraints to cover any types of errors. In this work, we demonstrated that DQM or \textit{SWITCH} performs much better than any feasible baselines. Note that without the ground truth, there are not many ways to work with possibly fallible data cleaning approaches (e.g., algorithm or crowd workers). 
However, one question still remains: How much trust can an analyst place in our estimates? 

We believe that the \textit{SWITCH} estimator actually provides valuable information, in addition to the currently observed crowdsourced cleaning results (\textit{VOTING}), to assess the quality of data when the ground truth is not available. However, it should be noted that \textit{SWITCH} is not able to estimate very hard-to-detect errors (i.e., black swan events) and assumes that the workers are  better than a random guesser, which holds true in practice \cite{DBLP:conf/sigmod/FranklinKKRX11,DBLP:conf/nips/LiuPI12,DBLP:conf/nips/ZhangCZJ14}.

%% file: related.tex
\section{Related Work}
To the best of our knowledge, this is the first work to consider species estimation for data quality quantification. 
Species estimation has been studied in prior work for  distinct count estimation and crowdsourcing~\cite{DBLP:conf/vldb/HaasNSS95, trushkowsky2013crowdsourced,unknownunknowns}.
However, the previous work only considered species estimation on clean data without false positive and false negative errors, which is inherent to the data quality estimation setting.
There are also several other relevant lines of work related to this problem:

\noindent\textbf{Label Estimation In Crowdsourcing: } The problem of estimating true labels given noisy crowds is well-studied.
For example, CrowdDB used majority votes to mitigate the effects of worker errors~\cite{DBLP:conf/sigmod/FranklinKKRX11}.
There have also been several proposed statistical techniques for accounting for biases between workers~\cite{DBLP:conf/nips/LiuPI12,DBLP:conf/nips/ZhangCZJ14}.
Other approaches leverage gold-standard data to evaluate the differences between workers~\cite{oleson2011programmatic}.
While the work in this area is extensive, all of the prior work studies the problem of recovering the true values in the already cleaned data.
In contrast, we explore the problem of estimating the number of errors that may be missed in the uncleaned or sparsely cleaned dataset.

\noindent\textbf{Progressive Data Cleaning: } When data cleaning is expensive, it is desirable to apply it \textbf{progressively}, where analysts can inspect early results with only $k \ll N$ records cleaned.
Progressive data cleaning is a well studied problem especially in the context of entity resolution \cite{altowim2014progressive, whang2014incremental, papenbrock2015progressive, gruenheid2014incremental}.
Similarly, sampling is now a widely studied in the context of data cleaning~\cite{wang1999sample, heise2014estimating}.
A progressive cleaning approach would assume a perfectly clean sample of data, and infer rules from the sample of data to apply to the rest of the dataset.
This is similar to the extrapolation baseline evaluated in this paper.
These techniques often assume that their underlying data cleaning primitives are infallible (or at least unbiased).
When these techniques are integrated with possibly fallible crowds, then we arrive at the species estimation problems described in this paper.

\noindent\textbf{Data Quality Metrics: } There have also been a number of different proposals for both quantitative and qualitative data quality metrics~\cite{DBLP:journals/cacm/PipinoLW02, DBLP:journals/jdiq/CheahP15, DBLP:journals/jdiq/EvenS09,DBLP:journals/jdiq/SessionsV09, DBLP:journals/tkde/FanGLX11,DBLP:journals/sigmetrics/KeetonMW09,yakout2011guided,bohannon2005cost,chomicki2005minimal,cong2007improving,lopatenko2007efficient}.
Most of the techniques rely on assessing the number of violated constraints or tests designed by the user, or qualitatively measure the number of erroneous decisions made by programs using the data; the major drawback of these techniques is that they work on the basis of having the ground truth. On the other hand, our statistical approaches are designed to work without the ground truth for any/mixed types of errors that are countable.
In terms of probabilistic techniques, Sessions et al.~\cite{DBLP:journals/jdiq/SessionsV09} learn a Bayesian Network model to represent the data and measured how well the model fits the data.
This formulation is not sufficient for our problem where we consider large number of missing data.
We explore a statistical formalism for data quality assessment (during progressive cleaning) when the data cleaning algorithms are fallible, and there is no available ground truth.

%% file: conclusion.tex
\vspace{-2mm}
\section{Conclusion}
\vspace{-0.5mm}
In this paper, we explored how to quantify the number of errors that remain in a data set after crowd-sourced data cleaning. 
We formalized a surprising link between this problem and the species estimation problem. 
However, we found that the direct application of species estimation techniques is highly sensitive to false positives during cleaning, and we proposed an alternative estimator, \textit{SWITCH}, to address the issue. 
Our experimental results suggest that this new estimator, which is based on switches in worker consensus on records, is more robust to both false positives and false negatives. 

We believe that variants of our approach could also apply to pure algorithmic cleaning (e.g., various machine learned error classifiers \cite{getoor2012entity}).
That is, instead of semi-independent workers, one could use several semi-independent automatic algorithms.
One challenge will be to relax the assumption that workers are drawn from a single infinite population (i.e., workers are identical and consistent in their skill levels, with some noise).
The finite-sample species estimation problem is discussed in~\cite{DBLP:conf/vldb/HaasNSS95}, and we are interested in exploring such approaches in the future.